# Space-based Missile Defense

David Wright

Laboratory for Nuclear Security and Policy, MIT

Submitted to *Science and Global Security*
January 27, 2026

This paper reviews the technical issues underlying space-based boost-phase missile defense and examines the current technology available for space-based interceptors and the characteristics of the missiles such a system would face. It then analyzes a particular space-based missile defense system that has been proposed to intercept in boost, ascent, and midcourse phases to illustrate the details of such an analysis and the constraints imposed on such systems by the physics of operating in space.

The Trump administration's January 2025 Executive Order called for the development of a missile defense system (now called Golden Dome) that includes a space-based system of interceptor-satellites "capable of boost-phase intercept".[1] The requirement for intercepting during the boost phase of attacking missiles—while their rocket motors are burning and before any decoys have been released—is to avoid the problem of decoys that plague midcourse missile defenses.

Deploying orbiting interceptors, however, introduces a range of complications due to the physics of orbital mechanics, including the large number of interceptors needed to give continuous defense coverage and the large mass of interceptors that can maneuver quickly enough toward an intercept point.

This paper discusses current technology available for space-based interceptors, the requirements on the system imposed by current missile threats, and the key equations needed to analyze such a system.[2] It then applies this information to analyze a specific proposal for space-based defense to illustrate the key technical issues that arise in attempting to build a reliable and effective space-based defense system.

In particular, the Brilliant Swarms system proposed by Booz Allen Hamilton and the press coverage it generated raised questions about whether current technology makes such systems more feasible than in the past.[3] Past plans for space-based boost-phase defense include the Brilliant Pebbles system envisioned for the Strategic Defense Initiative (SDI) and Global Protection Against Limited Strikes (GPALS) systems but that was never built.[4]

While many companies appear interested in developing boost-phase interceptors, few details are available about them. Enough information has been released about the initial Brilliant Swarms concept, however, to allow a technical analysis. This paper assesses the Brilliant Swarms proposal and provides the tools to update this analysis as new information emerges about this system and systems being planned by other companies.

We assess the capabilities of Brilliant Swarms based on information from the Booz Allen website and a press conference it held on March 27, 2025.[5] During its press conference and on its website, Booz Allen stated that the Brilliant Swarms system is intended to intercept ballistic missiles during boost, ascent, and early midcourse phases of flight.

Of these phases, it argues that the real value Brilliant Swarms adds to current missile defense capabilities is its ability to intercept during boost and ascent phase. Its stated goal is to defeat ballistic missiles before "countermeasures, decoys, or multiple warheads can be deployed," which by its definition is before the end of ascent phase. It emphasizes this goal by saying that current long-range US defenses, which attempt to intercept in midcourse, are not designed to counter missiles with "sophisticated countermeasures, decoys, and re-entry vehicles," and that Brilliant Swarms would fill this "critical gap in the current MDS [Missile Defense System] architecture" by providing the capability to intercept in boost and ascent phases.[6]

We therefore analyze the ability of a system like Brilliant Swarms to intercept during boost and ascent phases. While many details of the Brilliant Swarms proposal are not publicly available, and many are likely to change with time, we can show that any system with the general characteristics of Brilliant Swarms, and particularly the low mass of its satellites, cannot deliver its promised capability.

The Booz Allen presenters stated that this system would cost $25 billion to develop and deploy and would use very lightweight satellites with masses of only 40 to 80 kilograms (kg) each, in orbits with altitudes of 300 to 600 kilometers (km).[7]

Booz Allen further describes Brilliant Swarms as a "tailorable system, utilizing a constellation of approximately 1,000 to 2,000 satellites" that would be able to "communicate and coordinate in real-time, acting as both threat sensors and hit-to-kill interceptors."[8] An interceptor-satellite is intended to destroy its target by running into it and destroying it by the force of the impact. For boost-phase intercepts, the interceptor attempts to destroy the missile booster while its engines are still burning, to keep it from getting the warheads up to the speed they need to reach their targets. The warheads themselves would likely remain intact but would fall short of their original targets.

Below we first provide background information about different types of missile defense systems and the characteristics of the attacking missiles that boost-phase defenses would face. We then analyze the features of the Brilliant Swarms interceptors in two possible constellations of 2,000 interceptors, and show what would be required for boost and ascent-phase intercepts in each case.

## Missile Defense Basics

### *Midcourse defense*

The current US defense against long-range missiles, the Ground-based Midcourse Defense (GMD) system based in Alaska and California, is designed to intercept warheads during the midcourse phase of flight, which has two advantages: This phase lasts for tens of minutes, and since it is at very high altitude the warheads follow simple ballistic trajectories under the influence of only gravity, allowing the defense to accurately determine their future positions as they are tracked by space and ground sensors.

On the other hand, midcourse defenses face a daunting problem, as noted above: During this phase the warhead travels in outer space, where there is no air resistance. Thus, lightweight objects follow the same trajectory as heavy warheads, which opens the door to a wide range of countermeasures, including decoys to confuse and overwhelm the defense.[9] Despite decades of work, the midcourse countermeasure problem remains an unsolved challenge.[10] This is true whether the midcourse interceptors are based on the ground or in space.

*Boost-phase defense*

The defense could avoid midcourse countermeasures by intercepting during the missile's boost phase, before it deploys decoys or multiple warheads. Boost phase also has the advantage that the rocket plume of the large missile booster provides a very bright object for the interceptor to home on. These advantages have led to US interest in boost-phase systems for decades.

The main difficulty of intercepting in boost phase is the very short time available. Because boost phase lasts only a few minutes, the interceptors must be positioned close to the early part of a missile's trajectory so they can reach it while the missile's rocket motors are burning. A global boost-phase defense would require placing interceptors in space, in low-altitude orbits that would be close to the boosting missile during the first few minutes of its trajectory.

Moreover, the missile's large, bright plume gives the interceptor only an approximate location of the body of the missile. As the interceptor homes on the missile, it still faces the difficult problem of locating the missile body accurately enough to collide with it. In addition, an interceptor hitting the missile booster would reduce the range of the missile but is unlikely to destroy the warhead, which could still explode wherever it lands.[11]

Basing boost-phase interceptors in space creates multiple problems. The first is that interceptor-satellites in low-altitude orbits move relative to the surface of the Earth, both because they are traveling in their orbits and the Earth is rotating under their orbits.[12] An interceptor that at a particular moment would be near enough to the trajectory of a given missile to intercept it would quickly move out of position. Another interceptor must be in place to move into position near that trajectory so that there are no gaps in coverage. This requires a very large number of interceptors in orbit to intercept even a small number of missiles launched from a relatively small area on the ground, such as North Korea.

Since even a large constellation would have only a few interceptors in position to intercept launches from a given location at a given time, an attacker could therefore overwhelm the defense by launching a salvo attack—launching multiple missiles nearly simultaneously from the same region on Earth. Moreover, since the interceptors are in low-altitude orbits, the attacker could use inexpensive ground-based interceptors to destroy some of the satellites before they pass over a given launch site.[13]

An additional problem is that since boost phase lasts only a few minutes, an interceptor must be able to quickly change its trajectory to leave its orbit and reach the boosting missile. The interceptor must therefore include a rocket booster large enough to change its velocity by a large amount, which significantly increases the mass of the interceptor, as discussed below.

In addition, because the missile's rocket motors are burning during boost phase, the missile is an accelerating and potentially maneuvering target. This means that the missile can change its trajectory during boost phase, so that the defense does not know accurately where the missile will be in the future. The defense must track the missile for long enough to estimate an intercept point along its future trajectory, and the interceptor must carry enough maneuvering propellant to correct its trajectory in flight when its sensors allow it to determine the actual intercept point. The missile's maneuvering also means that as the interceptor gets close to the missile, it must still have enough propellant to allow it to adequately maneuver to destroy the missile by physically running into it. Carrying propellant for these types of maneuvering also increases the mass of the interceptor.

*Ascent-phase defense*

Booz Allen uses the term ascent phase to refer to the short period following the end of boost phase when the missile releases its warhead (or warheads) and countermeasures. If the missile carries multiple warheads, it may have a small upper stage, or "bus," to release the warheads on different trajectories. The goal of ascent-phase defense is to intercept before the missile can release countermeasures and multiple warheads.

Attempting to intercept during this phase rather than during boost phase would give the interceptor somewhat more time to reach its target. However, the time constraints are still severe. In this case, the target may also be maneuvering, although not as much as during boost phase.

*Countermeasures to Boost and Ascent-phase Intercepts*

While a primary motivation for intercepting during boost and ascent phases is to avoid midcourse countermeasures, the missile can also use countermeasures to prevent boost and ascent intercepts. A key example is shortening the length of these phases by using solid-propellant missiles with short boost and ascent times, as discussed below.

In addition, evasive maneuvering by the missile can complicate the defense's ability to determine an intercept location and can exhaust the interceptor's maneuvering propellant. In addition, as discussed above, salvo launches can locally overwhelm the defense, and might include some inexpensive decoy boosters that do not carry nuclear warheads but which the defense could not distinguish from real targets by the time it needed to commit interceptors.

An additional problem is that interceptors in low-altitude orbits could be attacked in various ways. Inexpensive, short-range missiles carrying direct-ascent interceptors could destroy interceptors prior to the launch of long-range missiles, creating a hole in the defense. Since such a constellation of interceptors would be highly dependent on communication with sensors and between the interceptors, local jamming may also be highly disruptive. The vulnerability of satellite constellations is seen as an important issue that must be addressed.[14]

Other countermeasures that have been studied and/or developed by both the United States and Soviet Union include releasing rocket-propelled decoys and jammers during the later part of boost phase, replacing the final booster stage with multiple small booster stages that each carry one warhead or decoy package, and releasing the missile's warheads while its final stage is burning. The last method eliminates ascent phase altogether.[15]

These potential responses by an adversary mean that boost or ascent-phase defenses could not be relied on even if space-based interceptors appropriate to these phases could be built. In this analysis we consider only the question of what is required for the interceptor to reach the intercept point with the missile during boost or ascent phase. It is important to keep recognize, however, that countermeasures can prevent an interceptor from hitting the target even if it can reach it.

## The Target Missiles

Understanding boost and ascent-phase intercept requires first understanding the attributes of the missiles the defense will face, which will determine what is required of the interceptor-satellites.

The Booz Allen presenters stated that the boost and ascent phase of a long-range missile might generally last about five minutes, although they assume it might last seven to eight minutes for a North Korean missile.[16] They also stated that the defense would have a total of about 15 minutes to intercept in midcourse phase; this would correspond to roughly half the trajectory of a long-range missile since the flight time of a 10,000-km intercontinental-range missile would be about 35 minutes.

However, US and Russian long-range solid-propellant missiles burn out in about three minutes, and China's are believed to as well.[17] While older, liquid-propellant North Korean missiles such as the Hwasong-15 have a burntime of four to five minutes, its more recent Hwasong-18 uses solid propellant and appears to burn out in less than three minutes.[18] If Iran develops or acquires long-range missiles, they are expected to use solid-propellant.

Moreover, Defense Department studies have shown that if boost-phase defenses become a concern, long-range missiles can be designed with fast-burn boosters that burn out in less than a minute.[19]

In addition, ascent phase can be very short, especially for missiles carrying a single warhead and decoys, which would not require a bus that releases multiple warheads after burnout. In this case, the warhead's trajectory to the target is set by controlling the speed and direction of the missile booster at burnout, so the warhead and accompanying decoys can be released immediately after boost phase. A 2011 report by the Defense Science Board (DSB) looks at a dozen US and foreign missiles and shows that they can release objects within 10 to 20 s of burnout.[20]

Even if the missile carries several warheads, they can be released as Multiple Reentry Vehicles (MRVs), which are not targeted at separate individual locations and do not have to be released sequentially, as Multiple Independently-targetable Reentry Vehicles (MIRVs) do. MRVs instead spread over an area around the aim point due to variations in their reentry conditions, which can be useful in attacking large targets.

Missiles carrying MIRVs typically use a bus to release the warheads one at a time with a short maneuver in between to place them on different trajectories. This process has typically lasted up to several minutes, as shown in the DSB report, in part because there was no reason to shorten it. But there are known methods to reduce this time, such as using a small bus on each warhead so that targeting is not done sequentially. The new Russian Yars-M missile is reported to use this approach, replacing the final missile stage with individual propulsion units for each warhead, developed in part to complicate defenses.[21]

Moreover, as noted above, techniques have been developed to deploy warheads by the end of boost phase, eliminating ascent phase altogether.

The total distance a missile warhead can travel increases with its speed at burnout. Once burnout occurs, the warheads have reached the speed they need to travel the distance to the target. If an interceptor hits the MIRV bus carrying warheads aimed at the United States, they would still have enough speed to carry them to the US, although they would not necessarily land near their specific targets.

In his assessment of boost phase defenses, Ashton Carter wrote that "the value of attempting bus intercept is very unclear, and it usually does not figure prominently in [missile defense] discussions."[22]

These considerations mean that the realistic threat that a system like Brilliant Swarms must be able to defend against is a solid-propellant missile with at most a very short ascent phase.

For our analysis we assume that the boost phase of an attacking missile lasts for three minutes, and the ascent phase lasts for 30 s or less.

For a missile on a standard trajectory and having a burntime of three minutes, the boost phase of a 10,000-km range missile would end at an altitude of about 200 km, which is well below the 300 to 600 km altitude of the interceptors. If it then took an additional 30 seconds to deploy the warhead and decoys, ascent phase would end at about 300 km altitude (see Figure 1).

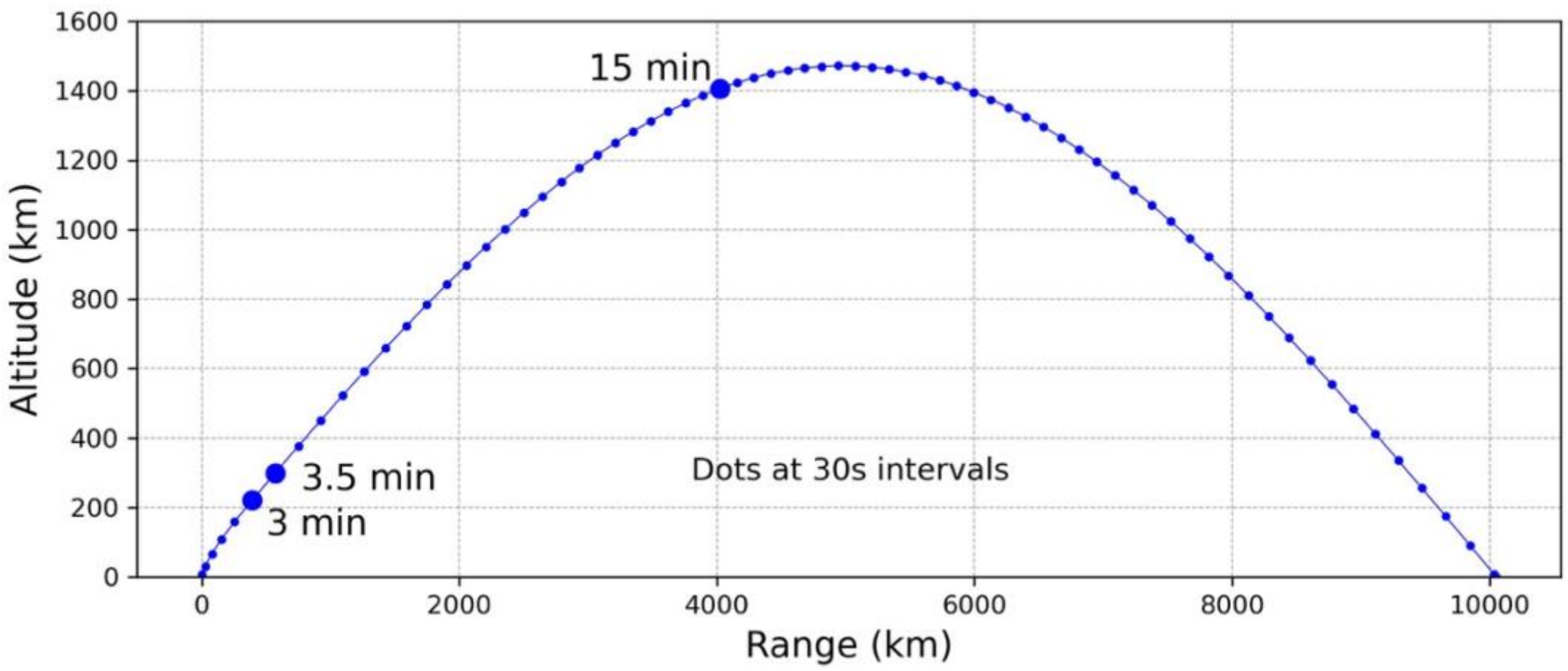

**Figure 1**. A standard trajectory for a 10,000 km-range missile. Booster burnout occurs at 3 minutes after launch and ascent phase is assumed to end by 3.5 minutes. The dots show the location at 30 second intervals.

### The Interceptor-Satellites

An interceptor-satellite consists of three parts:

(1) A kill vehicle that homes on and collides with the target missile or warhead. It carries sensors and a guidance system, as well as its own thrusters and propellant that it uses to maneuver as it approaches its target to collide with it.

(2) A rocket booster that accelerates the kill vehicle out of its position in orbit toward the estimated intercept point with its target. After the booster burns out, it separates from the kill vehicle. We refer to the combination of the kill vehicle and the booster as the "interceptor."

(3) A "lifejacket" that shields the interceptor from the space environment, and carries the solar panels that provide power as well as other equipment that the interceptor does not need to carry after it is fired (this is sometimes called the "host satellite" or "support satellite"). We refer to the combination of the interceptor and lifejacket as the "satellite." The lifejacket is left behind when the interceptor accelerates out of orbit toward an intercept point.

Under normal conditions, the interceptor-satellites move along predictable orbits. If the system detects a missile launch, it would determine which interceptor(s) would pass close enough to the missile's predicted trajectory to attempt an intercept.[23] That interceptor would then fire its booster to change its trajectory and move away from the position it would have if it remained in orbit, and fly toward the intercept point. The distance it can move from its point in orbit depends on the amount of propellant it carries to accelerate it and the time available to reach the intercept point.

We next estimate the interceptor-satellite mass required for boost or ascent-phase intercepts against solid-propellant missiles and compare that to the 40 to 80 kg satellite mass of Brilliant Swarms. As we show, most of the mass of an interceptor is due to the amount of propellant it must carry.

*Kill vehicle mass without propellant and tankage*

Booz Allen describes Brilliant Swarms as capable of intercepting targets during ascent and midcourse as well as boost phase. Since the target will not have a bright plume during ascent and midcourse phases, then in addition to the sensors needed for homing during boost phase, the kill vehicle sensor system must include long-wave infrared (LWIR) sensors similar to those of the Exo-atmospheric Kill Vehicle (EKV) used in the US GMD system. In order to detect a dim target at long distances, the EKV carries a sensor with a 20-centimeter aperture and an estimated mass of about 15 kg.[24] We assume that Brilliant Swarms could instead incorporate a version of the light-weight LWIR sensor used in the Clementine program but scaled up to have 20-cm optics and larger sensor arrays at two LWIR wavelengths (see Appendix A).

As noted, the kill vehicle must also carry a sensor that operates at wavelengths appropriate for boost-phase tracking, and likely a LIDAR system, which was also one of the Clementine sensors, for detecting the booster body while the sensor is homing on the booster plume.[25] We assume this full sensor package has a mass of 6.6 kg (see Appendix A). If the system was only designed to intercept in boost phase it would not need the LWIR sensors and the sensor package could be about 3 kg lighter.

The divert thrusters are the most massive components in the unfueled kill vehicle. We assume the kill vehicle uses a set of four thrusters in a cruciform pattern, as does the EKV and as is indicated in an illustration of a Brilliant Swarms interceptor on the Booz Allen website.[26]

Past studies determined that the thrusters must be able to produce 15 g's of acceleration in the last 10 seconds of homing in order to have sufficient thrust to intercept the booster.[27] That requirement rules out the use of very low-mass thrusters. We note in Appendix A that the lightest bi-propellant thrusters commercially available that can provide sufficient thrust appear to have a mass for four thrusters of about 13 kg. Including the other components required by the Divert and Attitude Control System (DACS) would further increase this mass. This system therefore appears to have a mass similar to, but somewhat larger than, the 14.5 kg assumed for the DACS in the 2003 APS study, which that study believed was not possible at the time but might be in the future. For our analysis, we will assume 14.5 kg as well.

The kill vehicle must also include an inertial measurement unit (IMU), an avionics package, communications equipment, batteries, and structures to hold these systems together during the high-acceleration maneuvering required for boost-phase intercepts.

For the calculations below we assume that the kill vehicle mass—excluding the kill vehicle's propellent and tankage—will be about 30 kg (see Appendix A).[28]

Moreover, since the goal of Golden Dome is to build defenses in the next few years, they must rely on existing, space-tested technology rather than possible future systems. For this reason, and to keep costs down, the presenters in Booz Allen's press conference discussed using off-the-shelf components for the satellites rather than high-end military components.

We note that attempting to design kill vehicles with lower mass than 30 kg would likely reduce the ability of the interceptor to home on and hit a target, and therefore reduce the effectiveness of the defense. For example, using smaller thrusters would give the vehicle less maneuverability to hit a boosting missile. As noted, a system designed only for boost-phase intercepts could be somewhat lighter since it would not need an LWIR sensor and the optics needed for homing on dim targets.[29]

*Kill vehicle mass with propellant and tankage*

The requirement to physically impact a boosting missile sets the amount of maneuvering the kill vehicle must be able to execute as it homes on the missile. Several technical analyses have determined that a space-based kill vehicle will require a change in velocity (its speed and/or direction) of 2.5 km/s for maneuvering to reliably hit a solid-propellant missile in boost phase—assuming the interceptor's booster has brought the kill vehicle close enough to its target.[30] Designing a kill vehicle with less than this divert capability would reduce its ability to hit and destroy missiles.[31]

Given the required velocity change, the mass of the kill vehicle's propellant and fuel tanks can be calculated using basic physics. The well-known rocket equation shows that the mass of propellant and fuel tanks required to produce a velocity change of 2.5 km/s is more than twice the mass of the rest of the kill vehicle (see Appendix and B). Assuming the mass of the unfueled kill vehicle is 30 kg, the total mass of the kill vehicle (including propellant and fuel tanks) for intercepting during boost phase will be about 95 kilograms.

Note that while the satellite industry has benefitted from the miniaturization of components, the kill vehicle size is constrained by some of its components. For example, the LWIR sensors must have optics large enough to collect sufficient infrared radiation to allow the interceptor to detect and home on a dim target after the end of boost phase. Reducing the size of the optics can cripple the interceptor's capabilities. Similarly, the kill vehicle mass could be reduced by limiting the size of the thrusters or the amount of propellant the kill vehicle has for maneuvering, but doing so would significantly reduce the capability of the defense.

*Interceptor booster mass*

A large fraction of the interceptor-satellite mass is that of the rocket booster that accelerates the interceptor out of its position in orbit toward the intercept point. The amount of propellant needed will depend on the number and arrangement of interceptors in the constellation, which determines how much velocity change is required, as discussed in detail below.

To deliver this velocity change, we assume that the interceptor booster could provide an average acceleration of 15 g's (147 m/s$^2$) to get the kill vehicle up to the required speed.[32] We note that too high an acceleration could lead to a very short burn time for the interceptor booster to reach the required speed. While it is burning, the booster can maneuver based on updated information about the missile's trajectory to point the kill vehicle in the direction of the updated intercept point. A short booster burn time could therefore require a larger amount of divert fuel for the kill vehicle to compensate for maneuvers by the attacking missile.[33]

For cases that require large velocity changes, we assume the interceptor booster will have two stages, which can reduce the mass. For this analysis we assume a one-stage booster for velocity changes less than 3 km/s and a two-stage booster for velocity changes greater than this. The choice of 3 km/s has little effect on our results.

*Life jacket mass*

Past studies have estimated that because of all the functions it must carry out, the lifejacket mass would be roughly 50 percent of the mass of the interceptor it carries.[34] For our estimates below, we assume a lower fractional mass—40 percent. This value may underestimate the lifejacket mass and therefore is generous to the defense.

**Brilliant Swarms System**

Booz Allen described the Brilliant Swarms system at its March 27, 2025 press conference.[35] Representatives of the company did not provide additional details in response to follow-up questions, saying that some of the specific details provided at the press conference may change.[36] However, the briefers provided enough information to allow an analysis of the potential capabilities of the system they described. One presenter at that event was Lt. Gen (ret.) Trey Obering, who headed the Missile Defense Agency from 2004 to 2008. He was formerly an executive vice-president and is now a senior advisor at Booz Allen.

As noted above, the presenters stated the Brilliant Swarms interceptor-satellites would have a mass of 40 to 80 kg, which is much smaller than the required mass estimated by other experts for boost-phase intercepts. The very small size of these interceptors is presented as an advantage of this system—for example, it would require fewer space launchers to lift the system into orbit.

The analysis above, however, shows that for a system intended to do what Brilliant Swarms is advertised to so, the fueled kill vehicle alone would have a mass greater than 80 kg, and the interceptor booster and lifejacket would significantly increase that mass.

This means that an interceptor a mass of 40 to 80 kg could not perform the stated goals of Brilliant Swarms. As a comparison, we estimate below the mass of an interceptor satellite that could potentially have this capability. Past proposals for space-based defenses found that interceptors would have masses of roughly 1,000 kg or more.

The mass of propellant in the interceptor booster will depend on the number and arrangement of interceptors in orbit. We refer to the Brilliant Swarms configuration described by the presenters at Booz Allen's press conference as the original Booz Allen constellation. It includes 2,000 satellites arranged with 100 satellites in each of 20 orbital planes, and with the satellites grouped in "flights" of five. Having five interceptors at each location would in principle allow the system to intercept several missiles launched from the same area at the same time, which is a recognition of the threat of salvo launches and attacks on the satellites. The satellites would orbit at altitudes of 300 to 600 km, and the presenters said they would be in sun-synchronous orbits, which are nearly polar.[37]

Because China has mobile missiles and Iran is building them, the United States must be able to intercept missiles launched from essentially anywhere in China and Iran, whose southern borders are at latitudes of 20 to 25 degrees. The defense must therefore provide full defensive coverage at latitudes above about 25 degrees. Such a defense would also provide full coverage against missiles launched from any higher latitude, which includes Russia and North Korea. China's large silo-based missile fields are near 40 degrees latitude, as are North Korea's missiles.[38]

We assume the satellites are in 300 km-altitude orbits so they are closest to the boosting missiles. The 20 groups of five in each orbit would then be separated by about 2,100 km along their orbit. At the equator, the orbits would be separated by about 1,050 km and their spacing would decrease at higher latitudes; at 25 degrees latitude, the orbits would be separated by about 950 km.

Figure 2 shows a section of a version of the Booz Allen configuration with 2,000 satellites in 300-km orbits. The vertical lines represent the orbits and each dot represents a flight of five interceptor satellites in its orbit. The circle represents the region around one flight of interceptors that they could defend, meaning that an interceptor could reach a missile passing through the circle. The size of the circle, which we call the "coverage area" of the interceptor, depends on the speed of the interceptor and the amount of time the interceptor has to reach the missile at the intercept point.

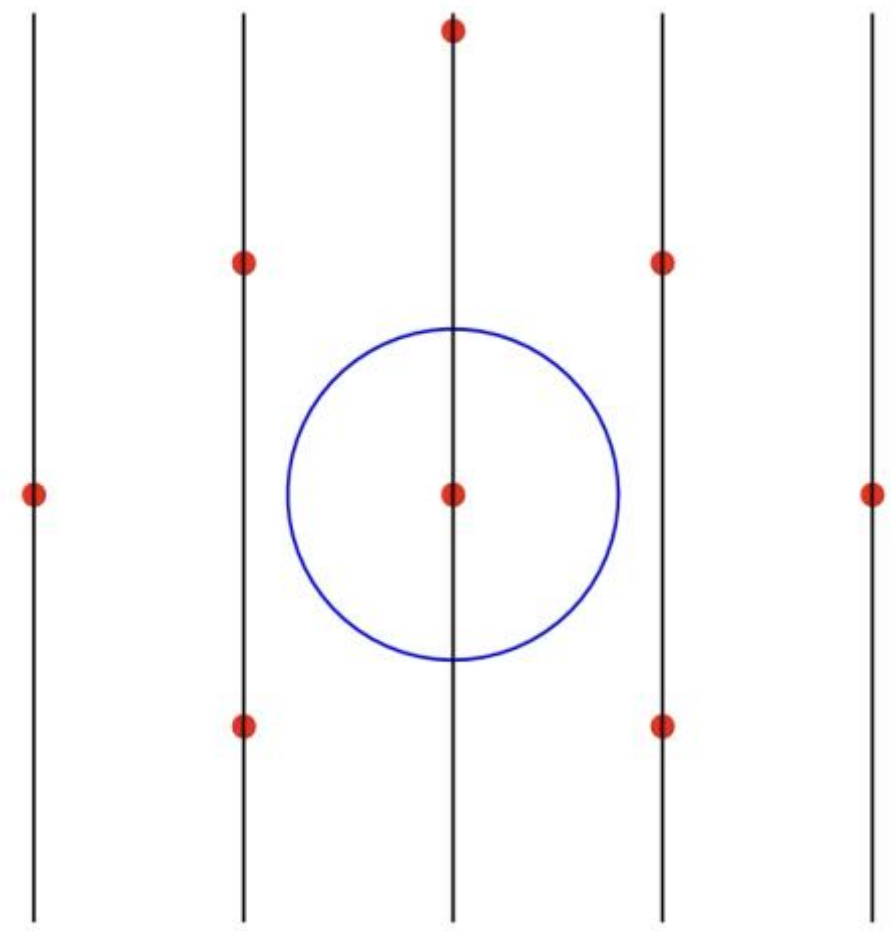

**Figure 2.** A version of the original Booz Allen Brilliant Swarms constellation, which includes 2,000 satellites in 20 orbits (indicated by the vertical lines) and 20 groups of five interceptors in each orbit (indicated by dots). The circle shows the coverage area of a single flight of interceptors at the dot in the center of the circle. For 300-km altitude orbits the groups of interceptors are separated by 2,100 km along the orbit, and the spacing between orbits in 1,050 km at the equator and 950 km at 25 degrees latitude.

Once the defense system detects the launch of a missile and tracks it long enough to determine its trajectory, it estimates an intercept point and calculates which interceptor will be closest to that point at the estimated time of intercept. That interceptor then fires its rocket booster and moves from its point in orbit, racing toward the estimated intercept point.

The amount of time available for the interceptor to reach the intercept point is actually less than the missile's boost or ascent time. Early warning satellites cannot reliably detect a solid-propellant missile for about 30 seconds after launch (until it reaches about 7 km altitude). Once the missile is detected the defense needs additional time to track the missile well enough to estimate its trajectory and an intercept point that the interceptor should fly toward.[39] The best current assessment is that developing a good estimate of an intercept point requires 15 seconds after detection of the missile.[40]

Roughly 45 seconds is therefore considered the earliest time the defense could calculate a "firing solution," which identifies the interceptor(s) that could reach the missile at the estimated intercept point and intercept time.

But the fact that the defense can fire the interceptor at this time does not mean that it should. Doing so would mean firing with no decision time after the firing solution is first determined. Between 30 and 45 seconds the missile will travel less than 10 km in altitude and range, providing the defense with very limited information on the type of missile and its ultimate trajectory. Since the defense has very few space-based interceptors in the right place to attempt an intercept, committing an interceptor this early would risk wasting it on a missile flying on a trajectory that would not pose a serious threat.

In addition, while there is interest in firing the interceptors as soon as possible, studies show that firing an interceptor too early would be counterproductive because there would still be large uncertainties in both the direction the missile is heading and the location of the intercept point. Compensating for these uncertainties would require additional maneuvering fuel for the interceptor that would further increase its mass.[41]

As a result, while presenters at the Booz Allen the press conference appeared to argue that "advanced tracking" using AI would allow Brilliant Swarms to fire interceptors very early, doing so is not likely to be an effective strategy, even if it is possible.

By 75 s after missile launch—which would allow 30 s of decision time—the missile would have traveled about 70 km and would be 40 percent through its burntime if it uses solid-propellant, placing considerably stronger constraints on the location of the intercept point. Moreover, at this point the first stage of a solid-propellant booster would have burned out, confirming whether the missile was liquid or solid-propellant, which is important for estimating its boost trajectory.

Below we calculate the results assuming both no decision time and 30 seconds of decision time, i.e., firing the interceptor at 45 and 75 s after missile launch.

In addition, for a boost-phase intercept to be effective, it must destroy the missile booster earlier than about 5 seconds before burnout in order to keep the booster from getting the warhead up to the speed it needs to reach its intended target.[42] This reduces the time available for boost-phase intercepts by an additional 5 seconds.

Finally, while we are assuming 180 seconds duration for boost phase, some sources report the burntime of North Korea's Hwasong-18 as 170 seconds, which would further reduce the time available for boost and ascent-phase intercepts.

**Requirements for Intercepting during Boost or Ascent Phase**

We next calculate the requirements for space-based interceptors in the Booz Allen constellation if they are to provide a defense against missiles during these two phases.

Figure 3 shows the coverage areas of several flights of five interceptors in the constellation shown in Figure 2, where in this case the radius of the coverage area for each flight of interceptors has been set to provide complete coverage against missiles launched at latitudes of 25 degrees and above. For complete coverage, the circles must be large enough that there are no gaps between them, given the spacing of the interceptors in the specified constellation; we refer to this as the "required coverage area" of an interceptor. In this configuration, some of the circles will overlap, which means that some areas will be covered by two sets of interceptors. Each satellite, and therefore each circular area, will travel along its orbit, and the interceptor must use its booster to move away from its orbital position at the dot to reach intercept points within the entire circular area, or above or below it.

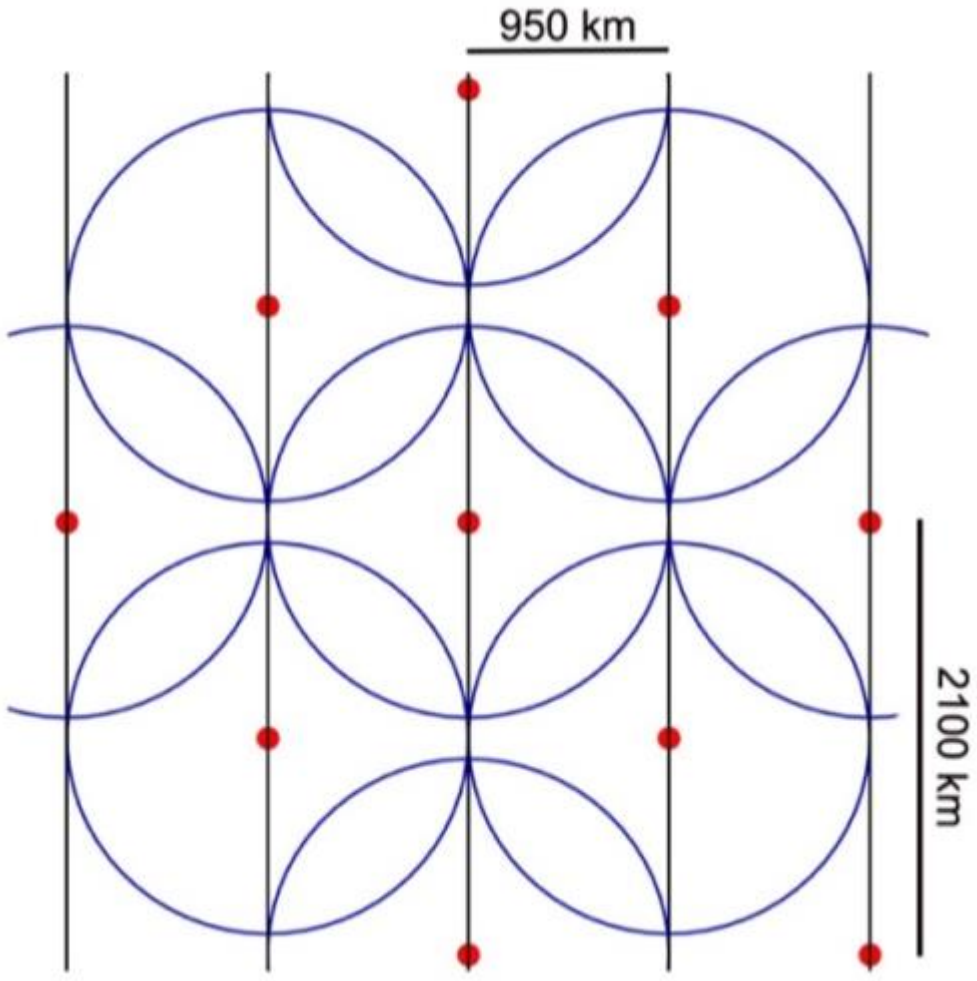

**Figure 3.** A version of the original Booz Allen Brilliant Swarms constellation shown in Figure 2, but with the circles showing the required coverage area of the interceptors to give complete defense coverage. The size of the circles is set so that there are no gaps in the coverage of the constellation. For 300-km altitude orbits the groups of interceptors are separated by 2,100 km along the orbit, and the spacing between orbits at 25 degrees latitude is 950 km. Each interceptor must be able to reach out in a circle of 950 km radius during the time it has to fly to intercept the missile, as well as the distance of the intercept point above or below 300 km.

At 25 degrees latitude, the orbits would be separated by about 950 km, which is then the radius of the required coverage area for each interceptor (see Figure 3). If a missile was heading toward the middle of the circular area, then the distance required to reach the intercept point would be less than 950 km, but complete coverage requires that an interceptor be able to travel about 950 km away from its orbital position during the boost/ascent time of the missile. In general, the interceptor will also need to fly a vertical distance above or below its orbital altitude to reach the missile at the time of intercept.

*Intercepting a solid-propellant missile during boost phase*

To intercept a 10,000 km range solid-propellant missile with a boost time of three minutes, and assuming the interceptor is fired with 30 seconds of decision time and impacts 5 s before burnout, an interceptor will have 100 seconds (3 minutes – 5 s – 75 s) to reach the intercept point. The missile will be at an altitude of 200 km at the end of boost phase, which is 100 km below the satellite orbit (see Figure 1). To intercept at the end of boost phase, the interceptor must travel 955 km—which is a little farther than the 950 km radius of its coverage area because it must also travel 100 km below its orbital altitude.[43]

As noted, we assume the interceptor booster can provide 15 g's of acceleration to move the interceptor out of its position in orbit and boost it to the required speed. However, even at this acceleration, the interceptor is unable to reach 955 km in 100 s (the equation for calculating the required velocity change with a finite acceleration is given in Appendix C).

In the case of no decision time, so that the interceptor was fired 45 s after the missile's launch, the interceptor would have to travel 955 km in 130 s, and the booster would have to provide 9.9 km/s of velocity change. We showed above that if the mass of the kill vehicle without propellant is 30 kg, then the fueled kill vehicle will have a mass of 95 kilograms. The rocket equation requires that to give a 95-kg kill vehicle a velocity change of 9.9 km/s, the interceptor would have a mass of more than 14 tons, giving a satellite mass of nearly 20 tons (see Table 1 and Appendix B).

We note that even if the missile burntime were 5 minutes, which might be appropriate to a North Korean liquid-propellant missile, the interceptor would have to travel 980 km since the missile would be about 250 km above the interceptors at burnout. This would require a velocity change of 4.2 to 4.8 km/s (for 0 and 30 s of decision time), leading to a satellite mass of 780 to 1030 kg.

Thus, an interceptor-satellite capable of conducting boost-phase intercepts throughout its required coverage area would be many times larger than the proposed Brilliant Swarms satellites.

How would this change if the interceptors only needed to intercept missiles by the end of ascent phase?

*Intercepting a solid-propellant missile during ascent phase*

Assuming ascent phase ends three and a half minutes after missile launch, if the interceptor were launched with 30 s of decision time it would have 135 seconds (3.5 minutes – 75 seconds)—rather than the 100 seconds for boost phase—to reach the intercept point. At three and a half minutes after launch, the missile's altitude—and thus the intercept altitude—will be about 300 km, which is the same as the orbital altitude (see Figure 1). The distance from the interceptor to the intercept point on the outer edge of the coverage area will be 950 km. The required velocity change of the interceptor in this case will be 9.1 km/s.

An interceptor booster required to give the 95-kg kill vehicle a velocity change of 9.1 km/s would give a total interceptor mass of 7,810 kg. Adding 40 percent of the interceptor mass for the lifejacket would give a satellite mass of 10,900 kg (See Table 1).

Assuming the interceptor is fired with no decision time, it would have 165 s to reach 950 km and would require a velocity change of 6.7 km/s, giving an interceptor mass of 1,870 kg, and a satellite mass of 2,620 kg.

| | **30 s decision time** | | | **0 s decision time** | | |
|---|---|---|---|---|---|---|
| | Velocity change (km/s) | Interceptor mass (kg) | Satellite mass (kg) | Velocity change (km/s) | Interceptor mass (kg) | Satellite mass (kg) |
| **Boost-phase intercept** | - | - | - | 9.9 | 14,200 | 19,900 |
| **Ascent-phase intercept** | 9.1 | 7,810 | 10,900 | 6.7 | 1,870 | 2,620 |

**Table 1:** Interceptor masses for the original Booz Allen constellation. To give full coverage, the interceptor must travel 955 km within 100 or 130 seconds for boost-phase intercepts (assuming 30 s and 0 s of decision time) and 950 km within 135 or 165 minutes for ascent-phase intercepts. For each case, the required velocity change by the interceptor booster and resulting interceptor mass are shown. The interceptor is assumed to provide 15 g's of acceleration. The satellite mass includes the interceptor and lifejacket, which is assumed to add 40 percent of the interceptor mass.

This analysis shows that to provide full coverage during ascent phase, interceptor-satellites in the original Booz-Allen constellation would, as for boost phase, require a much larger mass than the 80-kilogram proposal for Brilliant Swarms satellites.

## Capabilities of an Alternate Constellation

As noted, Booz Allen said that the details of the Brilliant Swarms constellation given at the March 2025 press conference might change. We therefore consider an alternative constellation, also with 2,000 satellites in 20 orbits at 300 km altitude, but with the satellites deployed uniformly along their orbits rather than in groups of five. In this case, the distance between the individual satellites within each orbit is 420 km rather than 2,100 km in the original constellation.

The constellation in Figure 4 appears to give the smallest required coverage area per interceptor for 2,000 interceptors in 20 near-polar orbits. The radius of the required coverage area of each interceptor is about 500 km, compared to 950 km for the original constellation. Below we estimate the mass of interceptors required to provide boost and ascent-phase defense throughout this coverage area.

Achieving the smaller interceptor spacing in Figure 4 with 2,000 interceptors requires that there is only one interceptor at each dot in Figure 4, rather than five interceptors in Figure 3. This arrangement would therefore reduce the number of interceptors at any location compared to the original constellation. It would only provide coverage by a single interceptor against a missile that passed through those regions where the circles do not overlap, and an attacker would know where those regions were by tracking the satellites. In regions where the circles overlap, the system could in principle provide coverage by two or three interceptors.

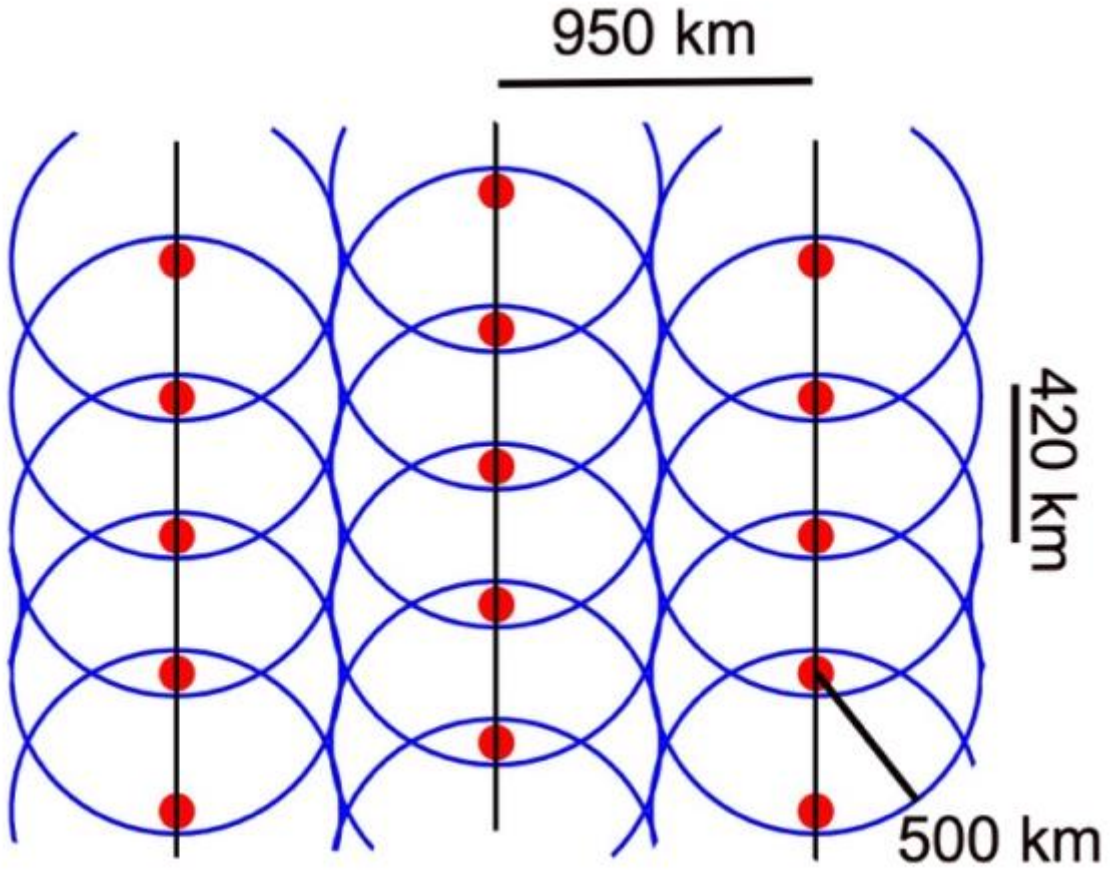

**Figure 4.** The alternate constellation. This constellation again has 2,000 satellites in 20 orbits (indicated by the vertical lines) with the same spacing between orbits as in Figure 3 (950 km at 25 degrees latitude). In this case the dots represent a single interceptor rather than a group of five as in Figure 3. The distance between the 100 interceptors within each orbit is 420 km for 300-km altitude orbits. For the interceptors to leave no gaps in coverage, each interceptor must be able to reach out in a circle of about 500 km radius during the time it has to fly to the target missile—compared to 950 km for the original constellation—as well as the distance of the intercept point above or below 300 km. In this constellation only a single interceptor could reach missiles passing through the areas where the circles do not overlap.

### *Intercepting a solid-propellant missile during boost phase*

We again assume a 10,000 km-range solid-propellant missile with a boost time of three minutes. As above, the interceptor will have a flight time of 100 seconds to reach the intercept point, assuming 30 s of decision time. Since the radius of its required coverage area is 500 km and the intercept point is 100 kilometers below the satellite orbit, the interceptor must be able to travel about 510 km in 100 seconds to reach any missile passing through its required coverage area.

If the interceptor booster provides an acceleration of 15 g's, this requires a velocity change of 6.6 km/s. A two-stage booster that could give a velocity change of 6.6 km/s to the 95-kg kill vehicle would have a mass of 1,680 kg, giving an interceptor mass of 1,775 kg and a satellite mass of 2,490 kg (see Table 2).

If the interceptor is fired with no decision time (45 s after missile launch), it would need to travel 510 km in 130 s. This would require a velocity change of 4.4 km/s, leading to an interceptor mass of 610 kg and a satellite mass of 855 kg.

| | **30 s decision time** | | | **0 s decision time** | | |
|---|---|---|---|---|---|---|
| | Velocity change (km/s) | Interceptor mass (kg) | Satellite mass (kg) | Velocity change (km/s) | Interceptor mass (kg) | Satellite mass (kg) |
| **Boost-phase intercept** | 6.6 | 1,775 | 2,490 | 4.4 | 610 | 855 |
| **Ascent-phase intercept** | 4.1 | 535 | 745 | 3.2 | 360 | 500 |

**Table 2:** Interceptor masses for the alternate constellation. To give full coverage, the interceptor must travel 510 km within 100 or 130 seconds for boost-phase intercepts (assuming 30 s and 0 s of decision time) and 500 km within 135 or 165 minutes for ascent-phase intercepts. For each case, the required velocity change by the interceptor booster and resulting interceptor mass are shown. The interceptor is assumed to provide 15 g's of acceleration. The satellite mass includes the interceptor and lifejacket, which is assumed to add 40 percent of the interceptor mass.

*Intercepting a solid-propellant missile during ascent phase*

If the goal is to intercept any missile in an interceptor's required coverage area within 3.5 minutes of the missile launch—by the end of ascent phase—the interceptor would have 135 s to reach the intercept point assuming 30 s of decision time. The distance to the intercept point with a missile passing through the edge of the coverage is 500 km, since it is at the same altitude as the satellite orbit. This case requires a velocity change of 4.1 km/s, resulting in an interceptor mass of 535 kg, and a satellite mass of 745 kg.

Even if interceptors were fired with no decision time and had 165 s to travel 500 km, the required velocity change would be 3.2 km/s, the interceptor mass would be 360 kg, and the satellite mass would be 500 kg.

As was the case for the original Booz Allen constellation, the satellite mass required to intercept in boost and ascent phases is much larger than the 40 to 80 kg proposed for Brilliant Swarms satellites.

Even if interceptors with the large masses calculated above were deployed, only one or two interceptors would be in place to attempt an intercept over most of the sky. Since a standard countermeasure to a space-based system would be to launch multiple missiles from the same area at the same time to locally overwhelm the defense, the defender would need to deploy multiple interceptor-satellites at each of the dots in Figure 4 to be able to provide a theoretically effective defense. Deploying five interceptors at each point, as Booz Allen proposes in its original constellation, to give a minimum of five interceptors in each part of space, would increase the total number of satellites in this constellation to 10,000.

**Defense coverage of a 240 kg satellite**

The analysis above illustrates the large mass required of interceptor-satellites that are able to provide boost-phase intercepts throughout their required coverage area. Would space-based interceptors smaller than these be a useful addition to US defense?

A less massive boost-phase interceptor than those discussed above would carry less propellant to accelerate the interceptor toward the intercept point, and therefore could attempt a boost-phase intercept throughout only a part of the area required for complete coverage.

As an example, consider an interceptor-satellite with a mass of 240 kg—three times the largest proposed Brilliant Swarms satellite. Assuming the lifejacket mass is 40 percent of the interceptor, that would give an interceptor mass of 171 kg. For a fueled kill vehicle with a mass of 95 kg, that would leave about 76 kg for the interceptor booster, which would be able to provide about 1.45 km/s of velocity change for the kill vehicle. Assuming 15 g's of acceleration, that velocity change would allow the kill vehicle to reach 140 km in the 100 s available for a boost-phase intercept attempt with 30 s of decision time, and a 180 km assuming no decision time. Since the interceptors must travel down 100 km to reach the boosting missile, that gives a radius of the coverage disc at the orbital altitude of 100 and 150 km for the cases of 30 s and no decision time, respectively

At mid-latitudes (45°), 240-kg interceptors that could reach a radius of 150 km could cover less than 5 percent of the sky for interceptors in the Booz Allen original constellation (see Appendix D).

Even in the alternate constellation these interceptors could cover less than a quarter of the sky at these latitudes (see Figure 5).

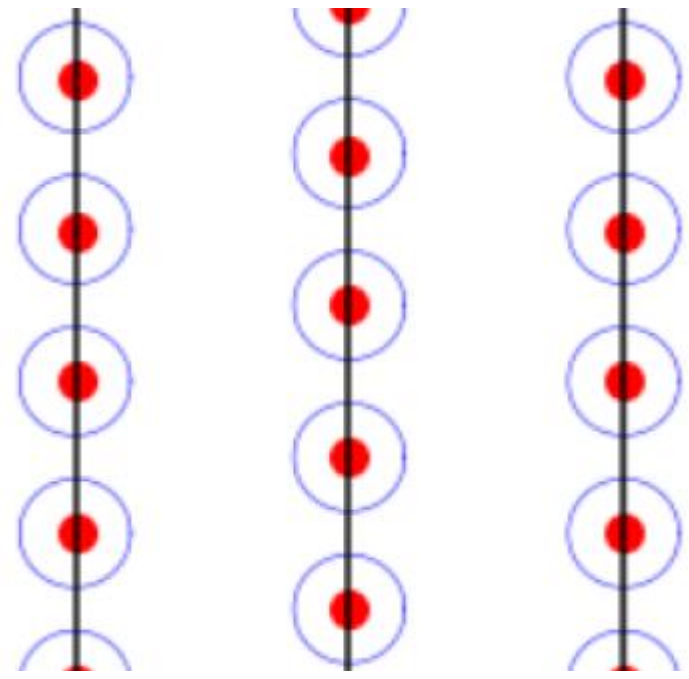

**Figure 5.** The circles show the boost-phase coverage provided by 240-kg space-based interceptors in the dense alternate constellation shown in Figure 4, at 45° latitude, assuming no decision time before firing the interceptors. Each dot represents a single interceptor in 300 km altitude orbits indicated by the vertical lines. In this constellation only a single interceptor could reach missiles passing through the areas within the circles. The covered area makes up only 23 percent of the sky at that latitude.

Leaving sections of the sky uncovered for boost and ascent-phase intercepts is essentially the same as providing no boost or ascent defense, for the following reason. The regions of sky that interceptors can cover will move as the satellites move in their orbits. However, satellites travel on predictable paths and by tracking the location of the interceptors an attacker would know what areas of sky were left undefended at any time and could launch missiles through the undefended areas. Any adversary capable of building a long-range missile would be able to do such tracking.

Closing the gaps in boost and/or ascent coverage for interceptors in the alternate constellation with a 150-km coverage radius would require about 14,000 interceptors—seven times as many as the proposed 2,000, yet over most of its coverage area the constellation could engage only one missile fired at a time from the same location.

**Midcourse Defense Capability**

The analysis above shows that satellites of the size proposed for Brilliant Swarms could not provide a boost or ascent-phase defense. An interceptor with such a small mass would have to be designed to carry only a small amount of propellant and could at best be used to attack objects during midcourse phase.

Such interceptors, however, could not provide effective defense in midcourse, either. As discussed above, during midcourse, interceptors face the problem of decoys and other countermeasures. Avoiding that problem by pursuing boost and ascent-phase defenses is the reason for putting interceptors in space in the first place and is the reason the White House announcement about Golden Dome specifically calls for boost-phase defenses.

Booz Allen states that Russian and Chinese missiles include "sophisticated countermeasures, decoys, and re-entry vehicles" and that US midcourse defenses were "not designed to counter these near-peer threats."[44] But putting midcourse interceptors in space does nothing to solve the countermeasures problem. In fact, if this system attempted to intercept early in a missile's midcourse trajectory, that would reduce the time that sensors in the system could observe the objects released by the missile, so that the system would have even less information that might be used for discrimination than would a ground-based midcourse defense system.

Moreover, the countermeasure problem is not limited to Russian and Chinese missiles. Any country capable of building a long-range missile and a warhead small enough to be carried by the missile—such as North Korea—would be capable of building effective countermeasures, and may be able to buy relevant technologies from other countries.[45]

**20-Year Cost of Brilliant Swarms**

In its 2025 press conference, Booz Allen stated it would cost $25 billion to deploy 2,000 Brilliant Swarms interceptors, including the cost of research and development of the system.[46] The presenters noted that the cost of building satellites that may be similar to Brilliant Swarms satellites is around $10 million per satellite, and that appears to underlie their overall cost figure.

All satellites have finite lifetimes, especially those in very low orbits since they exhaust the propellant they carry to counter the effects of orbital decay due to atmospheric drag. Booz Allen stated that the satellites would have a lifetime of four or five years, which means the entire constellation would have to be replaced every four or five years, leading to a large ongoing cost. Over 20 years, three or four new sets of satellites would be required, at a cost of about $20 billion for each set, assuming a cost of $10 million/per satellite.[47] As a result, the 20-year cost of a system like Brilliant Swarms would be $85 billion to $105 billion—not just the $25 billion initial cost.

**Conclusions**

The White House call for developing space-based boost-phase defenses has led to new interest in these systems. Intercepting missiles during their boost and ascent phases is seen as desirable since it could destroy the missiles before they released multiple warheads and countermeasures.

Given the characteristics of ballistic missiles in the arsenal of North Korea, as well as Russia and China, any defense of this kind must be designed to engage solid-propellant missiles with short burn and ascent phases, and that are equipped with midcourse countermeasures.

Our analysis illustrates the large amount of propellant space-based boost- and ascent-phase interceptors must carry to destroy missiles of this kind, which results in interceptors having very large masses—roughly half a ton or larger. It shows that proposals for systems intended to provide a global boost-phase defense at relatively low cost by using very lightweight satellites could not perform this mission.

In particular, our analysis finds that Brilliant Swarms and any system consisting of 1000-2000 low-mass satellites would provide essentially no defense capability against realistic missile threats during boost or ascent phase. Even if it could in principle engage boosting missiles in a small region around each interceptor-satellite, an attacking state could track the interceptors in orbit and calculate their future locations, and could therefore launch missiles through the undefended areas. For this reason, space-based defenses that offer partial coverage in practice offer none at all.

Our analysis shows that even using interceptor-satellites with masses of 240 kg—three times the mass of the proposed Brilliant Swarms interceptors—a constellation of 2,000 would leave very large areas of the sky with no defense coverage for boost or ascent intercepts. As a result, the system would provide no defense against a missile in boost or ascent phase.

Thus, while Brilliant Swarms is presented as a boost-phase capability in keeping with White House requirements for Golden Dome, a system with low-mass interceptors would instead be a *de facto* midcourse defense system. Building a midcourse defense, however, undercuts the rationale for putting interceptors in space in the first place, which is to attempt to intercept during boost or ascent phase to avoid the problem of decoys and other countermeasures that any midcourse defense would face. Countermeasures remain an unsolved vulnerability of midcourse defenses, whether based on the ground or in space.

For these reasons, a system like Brilliant Swarms would not be effective as a defense against missiles during any phase of flight.

**Acknowledgements**

The author thanks Lisbeth Gronlund, Fred Lamb, and Angelo Minotti for helpful discussions.

---

[1] "The Iron Dome for America," The White House, 27 January 2025, https://www.whitehouse.gov/presidential-actions/2025/01/the-iron-dome-for-america/

[2] David Wright, "Basic Equations for Analyzing Space-based Ballistic Missile Defense," 2026, https://lnsp.mit.edu/s/Wright-Basic-eqns-of-space-based-MD.pdf

[3] Booz Allen, "Brilliant Swarms for Golden Dome," https://www.boozallen.com/insights/fast-tracking-results/brilliant-swarms-and-the-golden-dome.html (accessed 21 January 2026).

[4] Donald Baucom, "The Rise and Fall of Brilliant Pebbles," The Journal of Social, Political and Economic Studies, Vol. 29-2, Summer 2004, p. 143, https://highfrontier.org/oldarchive/Archive/hf/The%20Rise%20and%20Fall%20of%20Brilliant%20Pebbles%20-Baucom.pdf.

[5] Booz Allen Press Conference, 27 March 2025, https://edge.media-server.com/mmc/p/s6yshtkc/ (accessed 22 May 2025); Sandra Erwin and Jason Rainbow, "Booz Allen unveils 'Brilliant Swarms' satellite concept for missile defense," *Space News,* 27 March 2025, https://spacenews.com/booz-allen-unveils-brilliant-swarms-satellite-concept-for-missile-defense/.

[6] Booz Allen, "Brilliant Swarms."

[7] Booz Allen Press Conference; Erwin, "Booz Allen unveils."

[8] Personal communication from Michael Keebler, Lead Associate, Reputation & Impact, Booz Allen Hamilton, Inc., 10 June 2025.
[9] A.M. Sessler et al., *Countermeasures,* (Cambridge, MA: Union of Concerned Scientists and MIT Security Studies Program, April 2000), https://www.ucs.org/sites/default/files/2019-09/countermeasures.pdf.
[10] American Physical Society (APS), Panel on Public Affairs, *Strategic ballistic missile defense: Challenges to defending the U.S.*, 3 March 2025, https://www.aps.org/publications/reports/strategic-ballistic-missile-defense. A key US approach to discrimination was to build the Long-Range Discrimination Radar (LRDR) at Clear, Alaska. However, for cost reasons this radar was built to operate in S-band rather than X-band. X-band could provide about three-times better range resolution and is typically said to be needed for discrimination; see "The LRDR: (Not) The Best Discrimination Money Can Buy?" MostlyMissileDefense blog, 30 January 2019, https://mostlymissiledefense.com/2019/01/30/the-lrdr-not-the-best-discrimination-money-can-buy-january-30-2019/
[11] David K. Barton, *et al*., "Report of the American Physical Society (APS) Study Group on Boost-Phase Intercept Systems for National Missile Defense: Scientific and Technical Issues," *Reviews of Modern Physics* 76, no. S1, 2003, https://journals.aps.org/rmp/abstract/10.1103/RevModPhys.76.S1, p. S241.
[12] Video on space-based missile defense, Union of Concerned Scientists, 12 September 2018, https://www.ucs.org/resources/space-based-missile-defense-not-good-idea
[13] Barton, "Report of the APS," p. S146; National Research Council (NRC), *Making Sense of Ballistic Missile Defense,* (Washington, DC: The National Academies Press, 2012), https://nap.nationalacademies.org/catalog/13189/making-sense-of-ballistic-missile-defense-an-assessment-of-concepts, p. 69, 142.
[14] Space News, "The Role of Space-Based Interceptors in Golden Dome," panel discussion, 12 November, 2025, https://spacenews.com/live-event-the-role-of-space-based-interceptors-in-golden-dome/. One potential response discussed in this panel, which is adding maneuverability to the support satellites that carry the interceptors, is unlikely to be effective: it would be difficult for a large support satellite to outmaneuver a homing ASAT interceptor, which can be small and highly maneuverable.
[15] Barton, "Report of the APS", p. S146; NRC, *Making Sense*, p. 69, 142.
[16] Booz Allen press conference.
[17] Burntimes of the US Minuteman III stages can be found at astronautix.com; Russia's Start-1 satellite launcher uses the stages of the RT-2PM Topol ballistic missile (Steven J. Isakowitz, Joshua B. Hopkins, Joseph. P. Hopkins, Jr., *International Reference Guide to Space Launch Systems, Fourth Edition*, (Reston, VA: American Institute of Aeronautics and Astronautics, 2004), p. 463).
[18] APS, *Strategic ballistic missile defense,* uses a burntime of 170 s for the Hwasong-18.
[19] Ashton Carter, *Directed Energy Missile Defense in Space–A Background Paper* (Washington, D. C.: U.S. Congress, Office of Technology Assessment, OTA-BP-ISC-26, April 1984), https://www.princeton.edu/~ota/disk3/1984/8410/8410.PDF.
[20] Defense Science Board (DSB), *Task Force Report on Science and Technology Issues of Early Intercept Ballistic Missile Defense Feasibility*, September 2011, https://apps.dtic.mil/sti/pdfs/ADA552472.pdf.
[21] Hans Kristensen, Matt Korda, Eliana Johns, and Mackenzie Knight, "Russian Nuclear Weapons, 2025," *Bulletin of the Atomic Scientists,* 81:3, 208-37, https://doi.org/10.1080/00963402.2025.2494386; Dmitry Kornev, and Alexey Ramm, "В ракетном темпе: какие перспективы у российских стратегических сил." *Izvestiya*, January 5, 2021, https://iz.ru/1105415/dmitrii-kornev-aleksei-ramm/v-raketnom-tempe-kakie-perspektivy-u-rossiiskikh-strategicheskikh-sil; Dmitri Kornev, *Telegram*, May 15, 2023, https://t.me/militaryrussiaru/5673.
[22] Carter, *Directed Energy,* p. 8.
[23] If the coverage areas of neighboring satellites overlap, more than one satellite may be within reach of the intercept point.
[24] NRC, *Making Sense*, Figure 5.6. This report also assumes 20-cm optics for a space-based midcourse interceptor. Note that an interceptor intended only for boost-phase intercepts could use a smaller optical system because a boosting target would be very bright. This report assumes such an interceptor would use 10-cm optics with a mass of 10 kg or less. As a result, a boost-only interceptor could be less massive than an interceptor intended for all phases.
[25] Michael Shannon, "The Clementine Satellite," *Energy and Technology Review*, Lawrence Livermore National Lab, June 1994, https://www.llnl.gov/sites/www/files/2020-05/clementine-etr-jun-94.pdf, Lunar and Planetary Institute, "The Clementine Mission: Instuments," nd, https://www.lpi.usra.edu/lunar/missions/clementine/instruments/.
[26] Booz Allen, "Brilliant Swarms."
[27] Barton, "Report of the APS," p. S231, S248.

[28] Note that the lightweight kill vehicle in Barton, "Report of the APS," which they assumed might be available by the mid-2010s, had a mass of 37 kg and did not include a large midcourse seeker.
[29] Some observers have raised the possibility of lowering the mass and cost of interceptors by reducing the kill probability of individual interceptors but compensating by launching two at each target (Space News, "The Role"). This plan would double the required number of interceptors in space; in order not to increase the total cost and mass in orbit this would require cutting the mass and cost of the interceptors in half, which does not appear feasible without significantly reducing the capability of the interceptors.
[30] Barton, "Report of the APS," p. S235, NRC, *Making Sense,* p. 240. Congressional Budget Office (CBO), *Alternatives for Boost-Phase Missile Defense*, July 2004, https://www.cbo.gov/sites/default/files/108th-congress-2003-2004/reports/07-22-missiledefense.pdf, p. 52, notes that engineers from Lawrence Livermore National Laboratory also supported a value of 2.5 km/s. Barton, "Report of the APS", p. S235, states that a divert capability of 2.0 km/s is sufficient for fast surface- and air-based interceptors.
[31] NRC, *Making Sense,* requires 2.5 km/s for maneuvering of the kill vehicle for boost phase intercepts of solid-fuel missiles. It also considers the case of a hypothetical space-based system intended to intercept in boost and midcourse phases that has 2 km/s rather than 2.5 km/s for maneuvering. This decrease would reduce its effectiveness for boost-phase intercepts, and would only make sense if the system could be effective enough in midcourse intercepts to destroy the missiles that slipped through the boost layer. This implies that the defense has figured out how to make midcourse work against countermeasures, which is not the current situation. The decrease in boost effectiveness would therefore limit the system's overall effectiveness. Note that the report's boost-plus-midcourse interceptor with 2 km/s still has a mass of 1,800 kg; see NRC *Making Sense*, p. 238 and Table E-18, p. 240.
[32] An example of a rocket motor that could give accelerations on this scale for an interceptor with a mass of about 600 kg is the Star 26B engine; see Northrop Grumman, *Propulsion Products Catalog*, n.d., p. 84, https://cdn.northropgrumman.com/-/media/wp-content/uploads/NG-Propulsion-Products-Catalog.pdf?v=1.0.0.
[33] The penalties for using significantly higher acceleration is discussed in Barton, "Report of the APS," p. S117.
[34] Barton, "Report of the APS," p. S112; NRC, *Making Sense,* p. 239.
[35] Booz Allen press conference; Erwin, "Booz Allen unveils."
[36] Personal communication, Michael Keebler.
[37] A 300-km-altitude sun-synchronous orbit would have an inclination of 96.6 degrees; orbits passing over the poles have an inclination of 90 degrees.
[38] Ryan Chan, "China Map Shows Nuclear Missile Silo Locations," *Newsweek*, 23 December 2025, https://www.newsweek.com/china-news-map-shows-nuclear-missile-silo-locations-2003927.
[39] Barton, "Report of the APS," S22, S170; APS, *Strategic ballistic missile defense*, p. 33; NRC, *Making Sense*, p. 41
[40] Barton, "Report of the APS," S22, S170; APS, *Strategic ballistic missile defense*, p. 33.
[41] See NRC, *Making Sense*, p. 63 and footnote 23; Dean A. Wilkening, "Airborne Boost-Phase Ballistic Missile Defense," *Science and Global Security*, 12 (2004), p. 8, https://scienceandglobalsecurity.org/archive/sgs12wilkening.pdf.
[42] Terminating the thrust of a 10,000 km range solid-propellant missile 4 s before its burntime would still leave it with enough speed to reach a range of about 8,000 km, which is sufficient to reach the northwest US from North Korea. Terminating thrust 5 s early would reduce the range to 7,500 km. Reducing the range below 5,500 km, which is the distance from North Korea to Alaska, would require terminating the thrust about 12 s early.
[43] The distance to the intercept point will be the hypotenuse of a triangle with sides of 950 and 100 km, which has a length of 955 km.
[44] Booz Allen, "Brilliant Swarms."
[45] Sessler, *Countermeasures*; National Intelligence Council, "Foreign Missile Developments and the Ballistic Missile Threat to the United States Through 2015," September 1999, https://web.archive.org/web/20250207052249/https://www.dni.gov/files/documents/Foreign%20Missile%20Developments_1999.pdf.
[46] Booz Allen press conference; Erwin, "Booz Allen unveils."
[47] This cost is in line with that in Todd Harrison, *Build Your Own Golden Dome: A Framework for Understanding Costs, Choices, and Tradeoffs,* American Enterprise Institute, AEI Foreign and Defense Policy Working Paper 2025-20, 12 September 2025, https://www.aei.org/research-products/working-paper/build-your-own-golden-dome-a-framework-for-understanding-costs-choices-and-tradeoffs/, p. 36, Table 4.

## Appendix A: Estimated Kill Vehicle Mass

*Kill Vehicle Mass without Propellant and Tankage*

To estimate the mass of the kill vehicle (KV) without propellant and tankage we follow the mass breakdown in the 2003 American Physical Society (APS) report and give updated estimates of the component masses.[1]

The APS report estimated a dry KV mass without tankage of about 37 kg using technology existing in 2003. Using a notional Lawrence Livermore design presented to the panel and adapting it to their requirements for a boost-based interceptor, the APS authors estimated that a KV mass of about 37 kg might be achievable within about a decade of the 2003 date of the report. We note that this is smaller than the estimated 40 kg dry mass without tankage of the Exo-atmospheric Kill Vehicle (EKV) of the Ground-based Midcourse Defense (GMD) system, even though the EKV needs less capable divert thrusters since it is designed for midcourse intercepts and does not contain the sensors needed for boost-phase intercepts.

Here we estimate what KV mass may be achievable today.

Mass of the sensor suite

The sensor suite mass is based on the masses of the sensors flown on the Clementine mission, modified as described below.[2]

A KV intended to home on targets in ascent and midcourse phase will need LWIR sensors similar to those of the EKV of the GMD system. The EKV uses 20-cm optics with cooled arrays in two LWIR bands. The 2012 National Research Council (NRC) study suggests that an improved version of the EKV should instead use 30-cm optics to reliably acquire a target at 2,000 km range.[3] That report estimates a mass of 45 kg for a sensor with 30-cm optics, and about 15 kg for one with 20-cm optics.[4]

In contrast, the Clementine LWIR sensor, which was used to detect thermal radiation from the moon's surface, has 13.1-cm optics and a mass of 2.1 kg.[5] Assuming the optics account for half the sensor mass (see Figure A1), and scaling the optics to a diameter of 20 cm by multiplying half the sensor mass by $(20/13.1)^2$, gives a sensor mass of 3.5 kg. Note that this mass is considerably smaller than the 15 kg assumed in the NRC report for a sensor with 20-cm optics.

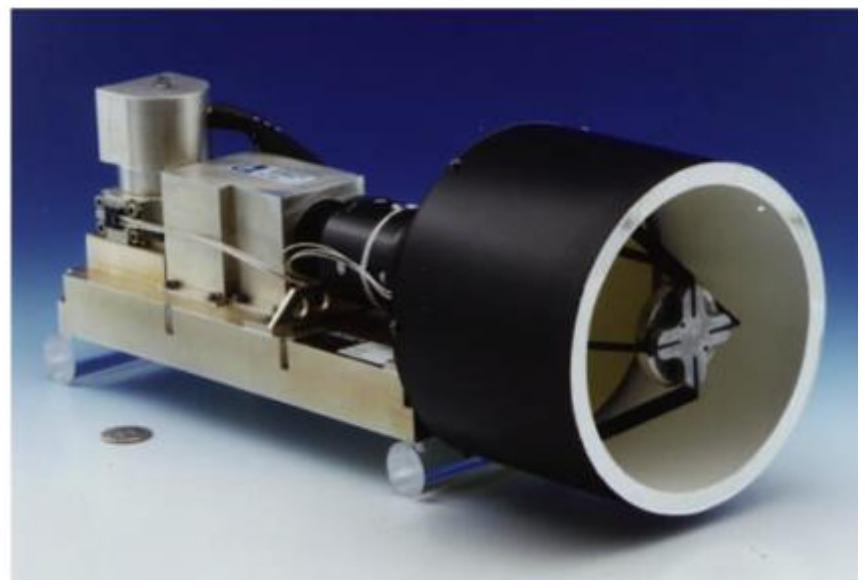

**Figure A1:** The Clementine LWIR sensor (Image from Shannon, "The Clementine Satellite").

The KV would also carry a cooled SWIR sensor for tracking the missile plume for boost-phase intercepts, a lidar system to locate the missile body as the KV homes on the missile plume, and an uncooled visible sensor. The masses of the corresponding Clementine sensors are:[6]

| | |
|---|---|
| SWIR (cooled): | 1.9 kg |
| Visible (uncooled): | 0.4 kg |
| Lidar: | 2.4 kg |

These four components give a total mass for the sensor suite of 8.2 kg, which is larger than the sensor mass in the APS report due to the larger LWIR sensor needed for post-boost intercepts.

Assuming there an overall 20 percent reduction in sensor mass in the additional decade beyond the estimates of the APS report, that would give roughly 6.6 kg for the sensor suite. (A KV intended only for boost-phase intercepts would not need the LWIR sensor, giving a sensor suite mass of about 3.8 kg.)

Divert and Altitude Control System (DACS) mass

As in the other studies, we assume the KV will maneuver using four thrusters in a cruciform arrangement, with each thruster able to produce 15 g's of acceleration during the last 10 s of homing (which is required for boost-phase intercepts) and capable of operating up to about 120 s.[7]

For a KV with dry mass of 30 kg, a 15 g acceleration would require each thruster to provide 4,400 N of thrust near the end of homing when most of the KV propellant has been used. The commercially available L3Harris AR-46 thruster can produce 7,000 N with a mass of 3.3 kg.[8] This recently developed thruster has the lowest mass for this thrust that we have identified commercially. Given the focus on developing low-mass components for commercial space uses, we assume this thruster is similar to what the military would use to build very large numbers of KVs for a constellation of interceptors.

A set of four of these thrusters would have a mass of about 13 kg plus additional mass for the fuel feed system. This total would likely be somewhat larger than the 14.5 kg assumed for the DACS in the 2003 APS report, so we use the smaller APS figure in our calculations.

Masses of other KV components

Assuming some advances in technology over the 2003 APS KV, we use the mass estimates given in Table A1 in our calculations.

| Component | This Study | APS 2003 Study |
|---|---|---|
| IMU | 0.5 kg | 1 kg |
| Avionics | 6 | 8 |
| KV Primary Battery | 1 | 1.8 |
| Structure | 2 | 3.5 |

**Table 2:** Masses of components assumed in this study compared to those used in the 2003 APS study.[9]

Combining these numbers gives an estimate of 30.6 kg for the KV mass without propellant and tankage. Given the uncertainties, for the calculations in this paper we round this off to 30 kg.

*KV Propellant and Tankage Mass*

Studies show that because of the high relative speed of interceptor and target missile for space-based boost phase interceptors, the KV requires a divert capability of $\Delta V = 2.5$ km/s.[10] The mass of propellant the KV must carry for this $\Delta V$ can be determined by the rocket equation, which is a statement of the conservation of momentum. This calculation is described in Appendix B. It assumes the tankage mass is 15 percent of the propellant mass, and that the effective exhaust velocity of the thruster is 2.9 km/s (corresponding to a vacuum specific impulse of about 310 s and an effective specific impulse of about 295 s).

In addition to the propellant mass calculated for the divert capability, we assume 2 percent of the propellant will remain as unusable fuel in the system, and that the attitude control system will require a 5 percent increase in the amount of propellant beyond that needed to provide $\Delta V = 2.5$ km/s.[11]

Assuming 30 kg for the KV mass without propellant and tankage, these assumptions lead to a fueled KV mass of about 95 kg (see Appendix B).

---

[1] David K. Barton, *et al*., "Report of the American Physical Society (APS) Study Group on Boost-Phase Intercept Systems for National Missile Defense: Scientific and Technical Issues," *Reviews of Modern Physics* 76, no. S1, 2003, https://journals.aps.org/rmp/abstract/10.1103/RevModPhys.76.S1, p. S253.
[2] Michael Shannon, "The Clementine Satellite," *Energy and Technology Review*, Lawrence Livermore National Lab, June 1994, https://www.llnl.gov/sites/www/files/2020-05/clementine-etr-jun-94.pdf.
[3] National Research Council (NRC), *Making Sense of Ballistic Missile Defense,* (Washington, DC: The National Academies Press, 2012), https://nap.nationalacademies.org/catalog/13189/making-sense-of-ballistic-missile-defense-an-assessment-of-concepts, p. 148.
[4] NRC, *Making Sense,* Figure 5-6, p. 151.
[5] NASA, "Clementine Long-Wavelength Infrared Camera (LWIR), 19 August 2011, http://nssdc.gsfc.nasa.gov/nmc/masterCatalog.do?sc=1994-004A&ex=03.
[6] Shannon, "The Clementine Satellite."
[7] Barton, "Report of the APS," p. S248.
[8] L3Harris, "Bipropellant Rocket Engines," 2025, https://www.l3harris.com/sites/default/files/2025-05/l3harris-ar-bipropellant-rocket-engines.pdf.
[9] Barton, "Report of the APS," S253.
[10] Barton, "Report of the APS," S235, NRC, *Making Sense,* p. 240. Congressional Budget Office (CBO), *Alternatives for Boost-Phase Missile Defense*, July 2004, https://www.cbo.gov/sites/default/files/108th-congress-2003-2004/reports/07-22-missiledefense.pdf, p. 52, notes that engineers from Lawrence Livermore National Laboratory also supported a value of 2.5 km/s. Barton, "Report of the APS", p. S235, states that a divert capability of 2.0 km/s is sufficient for fast surface- and air-based interceptors.
[11] Barton, "Report of the APS," p. S253.

**Appendix B: Adding Propellant and Tankage Mass to a Payload**

The rocket equation relates the initial ($M_i$) and final ($M_f = M_i - M_p$) masses of a rocket by:[1]

$$\frac{M_i}{M_f} = e^{\frac{\Delta V}{V_e}} \qquad (B1)$$

where $\Delta V$ is the velocity change from burning the propellant mass $M_p$; $V_e$ is the exhaust velocity of the thruster, given by $V_e = g*I_{sp}$, where $I_{sp}$ is the specific impulse of the thruster in units of $s^{-1}$ and $g$ is the gravitational acceleration.

Propellant and tankage are added to a payload mass $P$ to achieve a change in speed $\Delta V$. Some of the propellant will remain unburned; the useable mass of propellant is $M_p$, and the unburned amount is assumed to be a fraction $u$ of $M_p$. If the structural mass of the engine is $M_s$, then:

$$M_i = P + M_p(1+u) + M_s \qquad (B2)$$

$$M_f = P + uM_p + M_s \qquad (B3)$$

Assuming the tankage mass scales with the amount of propellant:

$$M_s = sM_p \qquad (B4)$$

then Eqs. B1 to B4 give:

$$M_i = P + (1+u+s)M_p \quad = \quad M_f e^{\frac{\Delta V}{V_e}} = \left(P + (u+s)M_p\right)e^{\frac{\Delta V}{V_e}} \qquad (B5)$$

Solving for the propellant mass gives:

$$M_p = P\,\frac{e^{(\Delta V/V_e)} - 1}{1 - r(e^{(\Delta V/V_e)} - 1)} \qquad (B6)$$

where $r = u + s$.

Using Equation B6, the factor by which the fuel and tankage increase the initial mass over the payload mass is:

$$\frac{M_i}{P} = \frac{P + (1+r)M_p}{P} = \frac{e^{(\Delta V/V_e)}}{1 - r(e^{(\Delta V/V_e)} - 1)}$$

$$= \frac{1}{(1+r)e^{-\Delta V/V_e} - r} \quad \equiv \quad F(r, \Delta V, V_e) \qquad (B7)$$

Structural mass estimates for tankage from technical reports on space-based interceptors give:[2]

> $s$ = 0.15-0.20 for the tankage of the kill vehicle's divert thrusters,
> $s$ = 0.1 for the larger rocket motor used to accelerate the kill vehicle out of orbit.

We assume $s = 0.15$, $u = 0.02$, and $V_e = 2.9$ km/s for the liquid-propellant thrusters of the kill vehicle. $V_e$ corresponds to a vacuum specific impulse of about 310 s, where the reduction takes into account the fact that the altitude control system (ACS) on the kill vehicle consumes fuel but does not contribute to the divert thrust.

We also assume that the ACS for the kill vehicle requires a 5% increase in the amount of propellant beyond that needed to provide $\Delta V = 2.5$ km/s.[3] The total propellant mass is therefore $(1 + w)M_p$, with $M_p$ given by Equation B6 and $w = 0.05$, and the fueled kill vehicle mass with ACS propellant is (ignoring the $rw$ term):

$$M_{KV}^{tot} = P + (1+r)(1+w)M_p \;=\; M_{KV}^{dry}\left[\frac{1+(1+w)\left(e^{\Delta V/V_e}-1\right)}{1-r(e^{\Delta V/V_e}-1)}\right] \qquad (B8)$$

where $P = M_{KV}^{dry}$ is the kill vehicle mass without propellant or tankage.

For the solid-propellant rocket motor of the interceptor, we assume $s = 0.1$, $u = 0$, and $V_e = 2.75$ km/s, which corresponds to an effective specific impulse of 281 s.

For a two-stage booster, the initial mass of the booster plus payload, assuming $V_e$ is the same for both stages, is given by:

$$M_i = P\,F(r_2, \Delta V_2, V_e)\,F(r_1, \Delta V_1, V_e) \qquad (B9)$$

where $\Delta V_1 + \Delta V_2 = \Delta V$ and $F$ is given by Equation B7. If $r_1 = r_2 = r$, $M_i$ is minimized for $\Delta V_1 = \Delta V_2 = \Delta V/2$, and this becomes:

$$M_i = P\,[F\left(r, \frac{\Delta V}{2}, V_e\right)]^2 \qquad (B10)$$

or

$$M_{int} = M_{KV}^{tot}\left[(1+r)e^{-\Delta V/2V_e} - r\right]^{-2} \qquad (B11)$$

where in this case $P = M_{KV}^{tot}$.

---

[1] David Wright, Laura Grego, and Lisbeth Gronlund, The Physics of Space Security (Cambridge, MA: American Academy of Arts and Sciences, 2005), https://www.ucsusa.org/resources/physics-space-security. This equation assumes a single stage booster.

[2] Barton, David K., *et al*., "Report of the American Physical Society Study Group on Boost-Phase Intercept Systems for National Missile Defense: Scientific and Technical Issues," *Reviews of Modern Physics* 76, no. S1, 2003, https://journals.aps.org/rmp/abstract/10.1103/RevModPhys.76.S1, p. 251; National Research Council (NRC), *Making Sense of Ballistic Missile Defense,* (Washington, DC: The National Academies Press, 2012), https://nap.nationalacademies.org/catalog/13189/making-sense-of-ballistic-missile-defense-an-assessment-of-concepts, p. 150.

[3] Barton, "Report of the APS," p. S253.

**Appendix C: Velocity Change Requirements for an Interceptor with Finite Acceleration**

If an interceptor must travel a distance *d* in a time *t* with an average acceleration *a*, what velocity change $\Delta V$ must its rocket motor provide?

Assume the rocket motor accelerates the interceptor at a constant rate *a* for a time $t_a$ to reach a speed $V_o$, and the interceptor then travels at $V_o$ until time *t*. In that case:

$$d = \frac{1}{2}a{t_a}^2 + V_0(t - t_a) \qquad (C1)$$

Since:

$$V_0 = at_a \qquad (C2)$$

Equation C1 can be written:

$$d = att_a - \frac{1}{2}a{t_a}^2 \qquad (C3)$$

Solving for $t_a$ and using Equation C2 gives:

$$V_0 = at\left(1 - \sqrt{1 - \frac{2d}{at^2}}\right) \qquad (C4)$$

Some of the velocity change provided by the interceptor booster will be used to change direction, but assuming that most is used to increase the interceptor's speed, then $\Delta V$ is essentially equal to $V_0$ and is given by Equation C4.

This can be rearranged to give the distance *d* the interceptor can travel in time *t*:

$$d = \Delta V\, t\left(1 - \frac{\Delta V}{2at}\right) \qquad (C5)$$

## Appendix D: Calculating the Fraction of Sky Covered by Space-based Interceptors

Consider a constellation of interceptor satellites, represented by the dots in Figure D1 below, in polar orbits that are separated by a distance *D*, with the interceptors in each orbit separated by a distance *d*. Each interceptor is assumed to be able to intercept missiles in the circle that surrounds it, which has a radius *r*.

We want to calculate the fraction of the sky that is contained within at least one circle, which is the fraction of the sky at that latitude that can be covered by these interceptors. Since the orbit separation *D* varies with latitude, this fraction depends on latitude and will not be the same globally.

Because the arrangement of circles repeats throughout the constellation, the fraction of the sky covered by the discs at this latitude can be found by calculating the fraction of the area in the rectangle with the dark outline in Figure D1 that is contained within at least on circle.

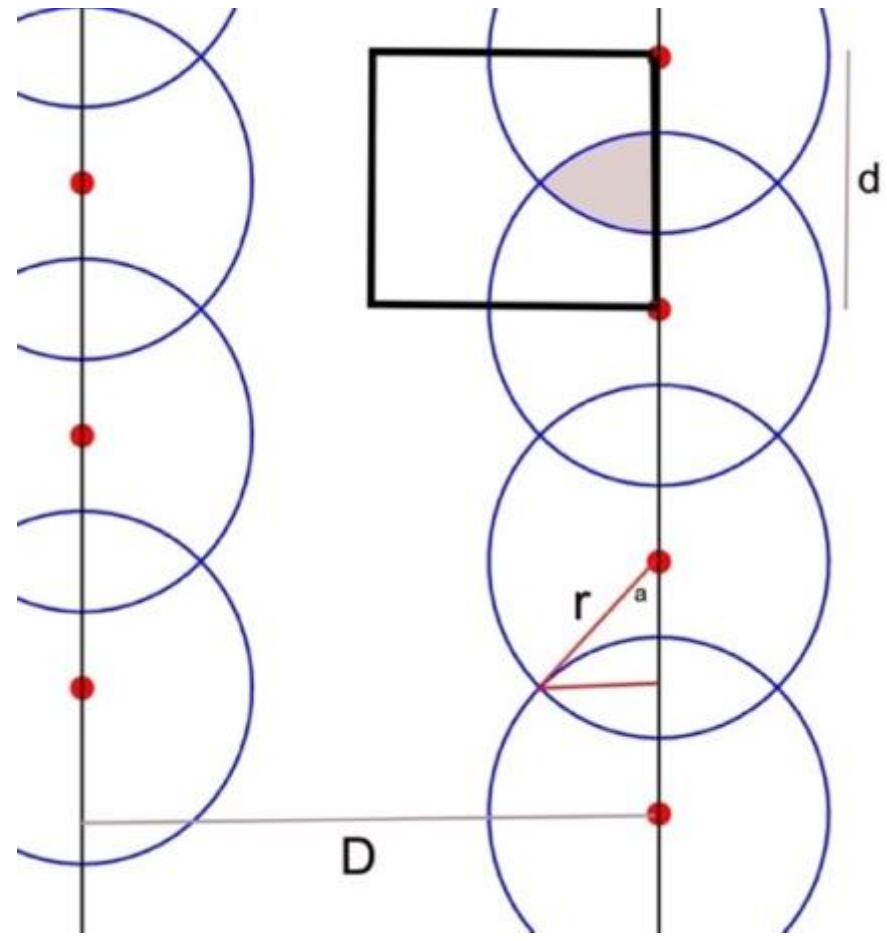

**Figure D1:** The dots represent interceptor satellites in the alternate constellation discussed in the text. The orbital planes are shown as vertical lines separated by a distance *D*, which depends on latitude. The interceptors are separated by a distance *d*. The circles of radius *r* represent the area around each interceptor that can be covered by that interceptor.

The area of the rectangle is *d***D*/2. Within the rectangle are two 90-degree wedge sections of the circles. Ignoring their overlap, those two wedges add up to half the area of a circle. The area of the rectangle that is within circles would then be half the area of a circle, except that the shaded area *A* is counted twice. Therefore, the fraction *f* of the sky at this latitude that lies within at least one circle is:

$$f = \frac{\frac{\pi r^2}{2} - A}{dD/2} = \frac{\pi r^2 - 2A}{dD} \quad (D1)$$

To calculate the area *A*, consider the triangle in the lower part of Figure D1. Half the area *A* can be calculated as the wedge swept out by moving the radius *r* through the angle *a,* minus the area of the triangle.

Therefore:

$$\frac{A}{2} = \left(\frac{a}{2\pi}\right)\pi r^2 - \frac{1}{2}\left(r\,sin(a)\right) * \left(r\,cos(a)\right) \quad (D2)$$

where the second term on the right is the area of the triangle. Since the vertical side of the triangle has the length $d/2$, the angle $a$ is given by:

$$a = \arccos\left(\frac{d}{2r}\right) \equiv \arccos(s) \quad (D3)$$

where:

$$s = \frac{d}{2r} \quad (D4)$$

Then:

$$A = ar^2 - \frac{d}{2}\sqrt{r^2 - \frac{d^2}{4}} = r^2\left(a - s\sqrt{1 - s^2}\right) \quad (D5)$$

This gives:

$$f = \frac{r^2}{dD}\left[\pi - 2\left(\arccos(s) - s\sqrt{1 - s^2}\right)\right] \quad (D6)$$

where $s$ is given by Equation D4.

In the case that $s > 1$, which means that $d > 2r$, the circles do not overlap, so that $A = 0$ and Equation D1 becomes:

$$f = \pi\frac{r^2}{dD} \quad (D7)$$

This derivation is valid as long as:

$$r \leq \frac{D}{2} \quad and \quad r \leq d \quad (D8)$$